\documentstyle[preprint,aps]{revtex}

\begin{document}

\title{The Nielsen Identities for the generalized $R_\xi$-gauge}
\author{C. Contreras$^1$\thanks{E-mail: ccontrer@newton.fis.utfsm.cl},
and L. Vergara$^2$\thanks{E-mail: lvergara@lauca.usach.cl}}
\address{$^1$ Departamento de F\'{\i}sica, Universidad T\'ecnica 
Federico Santa Maria \\ Valpara\'{\i}so,  Chile
\\$^2$ Departamento de F\'{\i}sica, Universidad de Santiago 
de Chile, Casilla 307,\\ Santiago 2, Chile}
\maketitle

\begin{abstract}
We show that it is possible, in opposition to a previous conjecture, to
derive a Nielsen identity for the effective action in the case of the
generalized $R_\xi $-gauge, where the gauge function explicitly depends on
the gauge parameter $\xi $. Also the Nielsen identity for the effective
potential is verified to one-loop in the Abelian Higgs model and the
corresponding identity for the physical Higgs mass is derived.
\end{abstract}


\section{Introduction}

The gauge dependence of the effective action $\Gamma (\varphi )$ and in
particular, that of the effective potencial $V_{eff}(\varphi )$ is known
since a long time ago \cite{Jackiw}. This fact raised the question about the
gauge invariance of the results obtained from the effective potential. For
example, it was not known whether spontaneous symmetry breaking was a gauge
invariant effect.

The investigation carried out in \cite{Nielsen} and \cite{FuKu} showed that
gauge independent physical quantities can be obtained from a gauge-dependent
effective potential and also that spontaneous symmetry breaking is a
gauge-invariant phenomenon.

In \cite{Nielsen} a set of identities was derived that implement the
physical requirement that the value of the effective potential at its
extrema should be left invariant under simultaneous variations of the gauge
parameter $\xi $ and the solution of the extremal equation. The works \cite
{Nielsen} and \cite{FuKu}, especially this last one, also allowed to
understand the origin of the problem studied in \cite{Weinberg} and further
analysed in \cite{DolJac}, on the definition of the effective potential in
the $R_\xi $-gauge.

In \cite{AITCH} the formal work done in \cite{Nielsen} was supplemented by
an explicit and complete calculation at the one-loop level in the Abelian
Higgs model. There the gauge-fixing function that defines the modified $%
R_\xi $-gauge was chosen as

\begin{equation}
\label{g.f.}F=-\partial _\mu A_\mu +ev\Phi _2, 
\end{equation}

\noindent where $v$ is an additional arbitrary gauge parameter that in
principle is not related to any property of the scalar field and $\Phi _2$
is the imaginary component of the scalar field.

Although the work done in \cite{AITCH} is very complete, they were unable to
prove that even in the simplest case the $\xi $-dependence of the effective
action $\Gamma $ could be expressed by means of a Nielsen identity in the
case of the generalized $R_\xi $-gauge where $v$ $=f(\xi )u$. In fact, they
conjectured that no dependence of $v$ on the gauge parameter is possible if
the Nielsen identities are to hold.

However, in a recent paper, \cite{Kobes} a generalization of the Nielsen
identities has been derived, which are also valid for the case of background
field gauges, in which the gauge fixing (\ref{g.f.}) is included.

In this letter we show explicitly, following the simple arguments shown in 
\cite{Kobes}, that it is possible to derive a Nielsen identity for the
effective action in the case of the generalized $R_\xi $-gauge where $v$
depends on the gauge parameter $\xi $, thus showing that the conjecture done
in \cite{AITCH} was not correct. The derivation is such that is readily
generalizable to non-abelian gauge theories.

For completeness we will perform an explicit calculation to one loop order
in the Abelian Higgs model in order to verify the Nielsen identity for the
effective potential. In addition we will show that the corresponding
identity for the physical Higgs mass is straightforwardly derived from the
identity for the effective action.

\section{Derivation of the Nielsen identity}

Let us consider a gauge theory in euclidean space defined by some Lagrangian 
${\cal L}(\phi_i)$, which includes a gauge-fixing Lagrangian given by

\begin{equation}
\label{lgf}{\cal L}_{g.f.}=\frac 1{2\xi }\left[ F(\phi _i;\xi )\right] ^2, 
\end{equation}
\noindent where $\phi _i$ denote the set of gauge and scalar fields. We
assume that with such a gauge-fixing function the Fadeev-Popov ghost fields
do not decouple. Therefore the generating functional of connnected Green
functions is

\begin{equation}
\label{gen}e^{-W(J,\xi )}=\int {\cal D}\phi _i{\cal D}\eta {\cal D}
\overline{
\eta }e^{-S-\frac 1{2\xi }\int d^dxF^2+\int d^dx\overline{\eta }(x)\frac{
\delta F}{\delta \phi _i(x)}\Delta _i\eta (x)+\int d^dxJ_i(x)\phi _i(x)}, 
\end{equation}

\noindent where $\Delta _i$ is defined via the BRS transformation of the
fields $\phi _i$ and $S$ is the classical action of the theory.

Using the fact that the integral over a single Grassman variable vanishes
and that the full action is invariant under the BRS transformations

\begin{equation}
\label{BRS}\delta \phi _i=\Delta _i\eta \omega ,\delta \overline{\eta }
=-\frac 1\xi F\omega ,\delta \eta =0, 
\end{equation}

\noindent it is easy to see that the following identity is satisfied

\begin{equation}
\label{ident1}\left\langle -\frac 1\xi F(\phi _i(x);\xi )K(\phi _i(x))+
\overline{\eta }(x)\frac{\delta K(\phi _i(x))}{\delta \phi _i(x)}\Delta
_i\eta (x)-\int d^dyJ_i(y)\Delta _i\eta (y)\overline{\eta }(x)K(\phi
_i(x))\right\rangle =0, 
\end{equation}

\noindent where use of eqs. (\ref{BRS}) has been done as well as the
properties of Grassman variables. Above, $K$ is an arbitrary functional of
the fields $\phi _i$ and the expectation value of the operator $O(\phi
_i,\eta ,\overline{\eta })$ is given by

\begin{equation}
\label{EXP}\langle O\rangle =e^{W(J,\xi )}\int {\cal D}\phi _i{\cal D}\eta 
{\cal D}\overline{\eta }\,O\,e^{-S_F+\int d^dxJ_i(x)\phi _i(x)} 
\end{equation}

\noindent and $S_F$ is the full action, including ghost and gauge fixing
terms.

Now we determine how the functional $W(J,\xi )$ varies under infinitesimal
changes in the gauge parameter $\xi $, i.e. when $\xi \rightarrow \xi
+\Delta \xi $. From eq. (\ref{gen}) we have that to leading order in $\Delta
\xi $

\begin{equation}
\label{var2}\Delta W=-\Delta \xi \int d^dx\left[ \left\langle \frac 1{2\xi
^2}F^2\right\rangle +\left\langle \frac F\xi \frac{\partial F}
{\partial \xi }
-\overline{\eta }(x)\frac \delta {\delta \phi _i(x)}\left( \frac{\partial F}
{\partial \xi }\right) \Delta _i\eta (x)\right\rangle \right] . 
\end{equation}

Using the identity (\ref{ident1}) with $K=\frac 1{2\xi }F$ we find that the
first term on the r.h.s. of eq. (\ref{var2}) equals

\begin{equation}
\label{id1}\frac 1{2\xi ^2}\left\langle F^2\right\rangle =-\frac 1{2\xi
}\int d^dyJ_i(y)\langle \Delta _i\eta (y)\overline{\eta }(x)F\rangle +\frac
1{2\xi }\langle \overline{\eta }(x)\frac{\delta F}{\delta \phi _i(x)}\Delta
_i\eta (x)\rangle 
\end{equation}

\noindent and if we choose $K$ to be equal to $\frac{\partial F}{\partial
\xi }$ the second term on the r.h.s. of eq. (\ref{var2}) can be written as

\begin{equation}
\label{id2}\left\langle \frac F\xi \frac{\partial F}{\partial \xi }-%
\overline{\eta }(x)\frac \delta {\delta \phi _i(x)}\left( \frac{\partial F}
{\partial \xi }\right) \Delta _i\eta (x)\right\rangle =\int
d^dyJ_i(y)\left\langle \Delta _i\eta (y)\overline{\eta }(x)\frac{\partial F}
{\partial \xi }\right\rangle . 
\end{equation}

Therefore, eq. (\ref{var2}) reads

\begin{equation}
\label{identity}\Delta W=\Delta \xi \int d^dx\int d^dyJ_i(y)\left\langle
\Delta _i\eta (y)\overline{\eta }(x)\left[ \frac F{2\xi }-\frac{\partial F}
{\partial \xi }\right] \right\rangle , 
\end{equation}

\noindent where we have dropped a constant term $\frac{\Delta \xi }{2\xi }$
arising from the second term on the r.h.s. of eq. (\ref{id1}).

The Nielsen identity for the effective action $\Gamma (\varphi _i)$ is
obtained by performing the Legendre transformation of eq. (\ref{identity}),
with the condition that the classical fields $\varphi _i$ are kept fixed
under variations of the gauge parameter $\xi $ \cite{FuKu}. Therefore it is
easy to see that in the limit when $\Delta \xi $ tends to zero such identity
reads

\begin{equation}
\label{nielsiden}\xi \frac{\partial \Gamma }{\partial \xi }=\int d^dx\int
d^dy\frac{\delta \Gamma }{\delta \varphi _i(y)}\left\langle \Delta _i\eta (y)
\overline{\eta }(x)\left[ \frac 12F-\xi \frac{\partial F}{\partial \xi }
\right] \right\rangle _\Gamma , 
\end{equation}

\noindent where in this case

\begin{equation}
\label{vev1}\left\langle O\right\rangle _\Gamma =e^\Gamma \int {\cal D}\phi
_i{\cal D}\eta {\cal D}\overline{\eta }\,O\,e^{-S_F+\int d^dx\frac{\delta
\Gamma }{\delta \varphi _i}(\phi _i-\varphi _i)} 
\end{equation}

\noindent and thus only one-particle irreducible Green function are taken
into account.

\section{Verification of the Nielsen identity to one loop order}

In this section we verify as an example that the Nielsen identity for the
one-loop effective potential is satisfied. To this end we consider the
Abelian Higgs model in four-diemsional euclidean space with Lagrangian

\begin{equation}
\label{lagr}{\cal L=}\frac 14F_{\mu \nu }F_{\mu \nu }+\left| D_\mu \Phi
\right| ^2-m^2\Phi ^{*}\Phi +\frac \lambda {3!}\left( \Phi ^{*}\Phi \right)
^2, 
\end{equation}

\noindent with $\Phi =[\Phi _1+i\Phi _2]/\sqrt{2}=[(H+\varphi )+iG]/\sqrt{2}$
and choose the gauge-fixing Lagrangian to be

\begin{equation}
\label{lgf1}{\cal L}_{g.f.}=\frac 1{2\xi }\left[ -\partial _\mu A_\mu +\xi
euG\right] ^2, 
\end{equation}

\noindent where for simplicity the simplest case of the generalized $R_\xi$
-gauge, $v$ $=\xi u$, has been considered. For this kind of gauge fixing
there is a mixing between the gauge and Goldstone degrees of freedom and the
ghosts fields do not decouple.

Assume a constant classical field $\varphi $ for the Higgs field component
and all other classical fields equal to zero. Therefore, from eq. (\ref
{nielsiden}) the Nielsen identity for the effective potential reads

\begin{equation}
\label{der1}\xi \frac{\partial V}{\partial \xi }=C\frac{\partial V}{\partial
\varphi}, 
\end{equation}

\noindent where $C$ is defined by

\begin{equation}
\label{c}C=-\frac 12\int d^dx\int d^dy\left\langle \left( \partial _\mu
A_\mu (x)+\xi euG(x)\right) eG(y)\eta (y)\overline{\eta }(x)\right\rangle . 
\end{equation}

\noindent The integrals have to be evaluated by dimensional regularization
and the infinite parts are discarded (the scale of dimensional
regularization will be set equal to one through out this letter). Thus,
expanding $V$ and $C$ in powers of $\hbar $

\begin{eqnarray}
\label{array}&& V=V^{(0)}+{\hbar}V^{(1)}+... \nonumber \\
&& C=C^{(0)}+{\hbar}C^{(1)}+... 
\end{eqnarray}

\noindent one finds that to one-loop order

\begin{equation}
\label{Nielspot}\xi \frac{\partial V^{(1)}}{\partial \xi }=C^{(1)}\frac{
\partial V^{(0)}}{\partial \varphi }, 
\end{equation}

\noindent where $C^{(1)}$ has a Feynman diagram representation shown in Fig.
1. The evaluation of $C^{(1)}$ requires the knowledge of the Goldstone,
ghost and mixed propagators, which are given by

\begin{equation}
\label{prop}\Delta_G=\frac{k^2+\xi m_A^2}{D_N};\;\;\;\;\Delta _{gh}=\frac{-1%
}{k^2+m_{gh}^2};\;\;\;\;\Delta _\mu =i\xi e(\varphi -u)\frac{k_\mu }{D_N} 
\end{equation}

\noindent respectively. Here,

\begin{eqnarray}
\label{array2}&& D_N=k^4+k^2(m_2^2+2\xi e^2\varphi u)+m_A^2[(\xi eu)^2+
\xi m_2^2],\nonumber \\
&& m_A^2=e^2\varphi ^2;\;\;\;\;\;m_2^2=-m^2+\frac \lambda 6\varphi
^2;\;\;\;\;\;m_{gh}^2=\xi e^2u\varphi,
\end{eqnarray} 

\noindent and in $\Delta _\mu $ the momentum flow is from $G$ to $A_\mu $.

\noindent Therefore, in momentum space, the sum of both terms in Fig. 1 gives

\begin{equation}
\label{Ce1}C^{(1)}=\frac{e^2\xi }2\int \frac{d^dk}{(2\pi )^d}\frac
1{D_N\,(k^2+m_{gh}^2)}\left\{ (\varphi -2u)k^2-\xi u\,e^2\varphi ^2)\right\}
. 
\end{equation}

Also, the $\xi $-dependent part of the one-loop contribution to the
effective potential is

\begin{equation}
\label{pot1}V^{(1)}=\frac 12\int \frac{d^dk}{(2\pi )^d}\ln D_N-\int \frac{
d^dk}{(2\pi )^d}\ln (k^2+m_{gh}^2) 
\end{equation}

\noindent and thus it derivative with respect to the gauge parameter reads

\begin{equation}
\label{devpot}\xi \frac{\partial V^{(1)}}{\partial \xi }=\frac
12(m_2^2\varphi )\xi e^2\int \frac{d^dk}{(2\pi )^d}\frac
1{D_N\,(k^2+m_{gh}^2)}\left\{ (\varphi -2u)k^2-\xi ue^2\varphi ^2)\right\}. 
\end{equation}

Now, combining

\begin{equation}
\label{pot0}\frac{\partial V^{(0)}}{\partial \varphi }=(-m^2+\frac \lambda
6\varphi ^2)\varphi =m_2^2\varphi 
\end{equation}

\noindent and eqs. (\ref{Ce1}) and (\ref{devpot}) we observe that the
Nielsen identity for the effective potential, eq. (\ref{Nielspot}) is
satisfied.

\section{Nielsen identity for the physical Higgs mass}

The physical Higgs mass $M^2$ is defined as the pole in the physical part of
the Higgs propagator (in Minkowski space). In the case of the abelian Higgs
model the inverse propagator is given by

\begin{equation}
\label{invprop}G^{-1}(x-y)=\frac{\delta ^2\Gamma }{\delta \Phi _1(x)\delta
\Phi _1(y)}\bigg\vert_{\Phi _1=\varphi } 
\end{equation}

\noindent and vanishes on the euclidean mass shell: $G^{-1}(p^2=-M^2)=0$.

Therefore, by differentiating the Nielsen identity for the effective action
twice with respect to $\Phi _1$, evaluating at $\Phi _1(x)=\varphi $ and the
using eq. (\ref{invprop}) we obtain

\begin{equation}
\label{green1}\xi \frac \partial {\partial \xi }G^{-1}(w-z)=\int d^dx\int
d^dy\frac \delta {\delta \varphi }G^{-1}(w-z)\left\langle \Delta _i\eta (y)
\overline{\eta }(x)\left[ \frac 12F-\xi \frac{\partial F}{\partial \xi }
\right] \right\rangle _\Gamma , 
\end{equation}

\noindent where as before all classical fields except that of the Higgs
field have been set equal to zero. This equation can be rewritten as

\begin{equation}
\label{green2}\xi \frac \partial {\partial \xi }G^{-1}(w-z)=C\frac \delta
{\delta \varphi }G^{-1}(w-z), 
\end{equation}

\noindent with $C$ defined in eq. (\ref{c}). Or, in Fourier space

\begin{equation}
\label{green3}\xi \frac \partial {\partial \xi }G^{-1}(p^2)=C\frac \delta
{\delta \varphi }G^{-1}(p^2). 
\end{equation}

\noindent This equation means that if the inverse propagator vanishes for
particular values of the momentum squared, the gauge parameter and $\varphi$
, it will also vanish for the same value of $p^2$if $\xi \rightarrow \xi
+\delta \xi $, $\varphi \rightarrow \varphi +C\delta \xi /\xi $. Therefore,
the Nielsen identity for the Higgs mass squared reads

\begin{equation}
\label{nielmas}\xi \frac \partial {\partial \xi }M^2=C\frac \delta {\delta
\varphi }M^2. 
\end{equation}

\noindent This equation can be directly verified to the (first nontrivial)
one-loop order and this will be done elsewhere.

\section{Conclusions}

Therefore we have seen that it is possible to obtain in a rather simple way
a Nielsen identity for the effective action in the generalized $R_\xi$
-gauge and verified that the corresponding identity for the effective
potential is satisfied to one-loop order, in scalar electrodynamics. This
result shows that the a previous conjecture on the impossibilty of deriving
such an identity is not correct. It is clear that although the verification
of the identity for the effective potential was done for the special case
$v$ $=f(\xi )u$, $f(\xi )=\xi $, our results are valid for an arbitrary, but
well behaved, function $f(\xi )$.

In addition, we have shown that the corresponding identity for the physical
Higgs mass can be straightforwardly obtained from eq. (\ref{nielsiden}).

Finally, it is clear that these ideas are easily generalizable to the case
of non-abelian gauge theories because in esence we have just made use of BRS
symmetries and the properties of Grassman variables.

\section{Aknowledgements}

This work was partially suported by a grant No. 049631VC of the Departamento
de Investigaciones Cient\'{\i}ficas y Tecnol\'ogicas (DICYT) of the
Universidad de Santiago de Chile.

\newpage
\section{Figure Captions}
Fig. 1. Feynman representation of $C^{(1)}$. The dashed line represents
the ghost propagator and the full line, the Goldstone propagator.
The mixed propagator is shown as a wiggly and full line. The crosses represent
$e$ times numerical factors.       

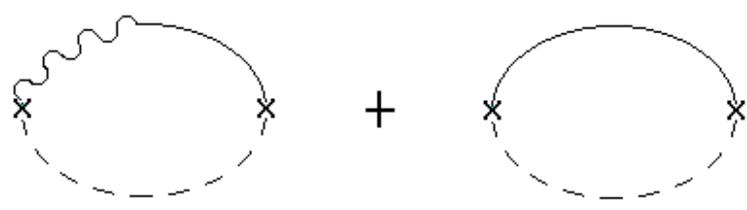
\end{document}